\documentclass[pdflatex,sn-mathphys-num]{sn-jnl}

\usepackage{graphicx}%
\usepackage{multirow}%
\usepackage{amsmath,amssymb,amsfonts}%
\usepackage{amsthm}%
\usepackage{mathrsfs}%
\usepackage[title]{appendix}%
\usepackage{xcolor}%
\usepackage{textcomp}%
\usepackage{manyfoot}%
\usepackage{booktabs}%
\usepackage{algorithm}%
\usepackage{algorithmicx}%
\usepackage{algpseudocode}%
\usepackage{listings}%

\theoremstyle{thmstyleone}%
\theoremstyle{thmstyletwo}%

\theoremstyle{thmstylethree}%

\begin{document}

\title[Article Title]{Screening in the Heitler-London Model: Revisiting the Bonding and Antibonding States of the Hydrogen Molecule}

\author[1]{Washington P. da Silva}
\equalcont{These authors contributed equally to this work.}

\author[1]{Daniel Vieira}
\equalcont{These authors contributed equally to this work.}

\author[2]{Jonas Maziero}
\equalcont{These authors contributed equally to this work.}

\author[1]{Edgard P. M. Amorim}\email{edgard.amorim@udesc.br}
\equalcont{These authors contributed equally to this work.}

\affil[1]{\orgdiv{Departamento de F\'isica}, \orgname{Universidade do Estado de Santa Catarina, 89219-710},\orgaddress{\city{ Joinville}, \state{SC}, \country{Brazil}}}

\affil[2]{\orgdiv{Department of Physics, Center for Natural and Exact Sciences}, \orgname{Federal University of Santa Maria, 97105-900}, \orgaddress{\city{Santa Maria}, \state{RS}, \country{Brazil}}}

\abstract{The present manuscript revisits one of the earliest approaches to treating molecular systems within the Schr\"odinger formalism of quantum mechanics: the Heitler-London (HL) model. Originally proposed in 1927 and based on a linear combination of atomic orbitals, the HL model provided a foundational description of covalent bonds and has served as the basis for numerous variational methods. Focusing on the hydrogen molecule, we begin by revisiting the analytical calculations of the original HL model, from which the qualitative physics of bonding and antibonding states can be obtained. Subsequently, we propose including electronic screening effects directly in the original HL wave function. We then compare our proposal with variational quantum Monte Carlo (VQMC) calculations, whose trial wave function allows us to optimize the electronic screening potential as a function of the inter-proton distance. We obtain the bond length, binding energy, and vibrational frequency of the H$_2$ molecule. Beyond revisiting this foundational approach in quantum mechanics, our proposal can serve as improved input for constructing new, but still analytically simple, variational wave functions to describe dissociation or bond formation.}




\maketitle

\section{Introduction \label{sec:1}}

Since the earliest stages of quantum mechanics, the hydrogen molecule (H$_2$)---the simplest neutral molecule---has been a central topic of investigation in both molecular physics and chemistry. Comprising two protons separated by a distance $R$ and surrounded by two electrons, the description of the hydrogen molecule still presents a complex challenge within the Schr\"odinger formalism. The first attempts to provide a full molecular picture date back to the beginning of the twentieth century. Given that exact analytical solutions are available for the hydrogen atom, a natural constraint arises: in the limit of large proton–proton separation, the molecular wave function of H$_2$ must reduce to a linear combination of single-electron hydrogen atomic orbitals, corresponding to two isolated hydrogen atoms without electronic interaction. Within this constraint, it was Heitler and London, in their seminal 1927 paper \cite{heitler1927wech}, who had the key idea of expressing the molecular wave function as a linear combination of products of atomic orbitals for any nuclear separation. Despite its simplicity, such a wave function provides a compelling quantum-mechanical description of the hydrogen molecule. Beyond satisfying the natural constraint for $R \to \infty$, it predicts the overlap of atomic orbitals for finite values of $R$, allowing the $1s$ electrons to be shared, which leads to the formation of bonding and antibonding molecular orbitals. The bonding molecular orbital has a lower energy than the combined energy of two separated hydrogen atoms. Consequently, a stable molecule is formed with a strong covalent bond. 

The calculation of the hydrogen molecule has been revisited many times over the years, with key developments marking the evolution of the HL model. In the late 1920s and early 1930s, Wang \cite{wang1928} investigated H$_2$ using what was then called ``new quantum mechanics'', while Hylleraas \cite{hylleraas1929} introduced correlated coordinates for helium, an idea extended to H$_2$ by James and Coolidge \cite{james1933}. These pioneering refinements highlighted the importance of electron–electron correlation beyond the original HL {\it ansatz}. From the 1960s onward, variational improvements played a crucial role. Kołos and Wolniewicz \cite{kolos1965,wolniewicz1966} introduced spheroidal coordinates with optimized variational parameters, while Cooley \cite{cooley1961} developed improved numerical schemes for solving the Schr\"odinger equation. Variational studies were also extended to hydrogen systems under strong magnetic fields, as in the works of Vincke and Baye \cite{vincke1985}, and later Doma \textit{et al.} \cite{doma2016}, further demonstrating the flexibility of HL-inspired approaches. With the progress of computational physics, high-precision studies became possible. Bishop and Cheung \cite{bishop1978}, Sims and Hagstrom \cite{sims2006}, Cencek and Szalewicz \cite{cencek2008}, and Pachucki \cite{pachucki2010,pachucki2013} carried out accurate calculations, later extended to include relativistic and quantum electrodynamical (QED) effects \cite{puchalski2023}. Such works established the HL model as a conceptual benchmark for comparing exact and approximate treatments. In parallel, stochastic and quantum simulation approaches have revitalized HL-based ideas in contemporary contexts. Quantum Monte Carlo studies (e.g., Chen and Anderson \cite{chen1995}, Corongiu and Clementi \cite{corongiu2009}, Prayitno \textit{et al.}~\cite{prayitno2023}) and quantum eigensolver demonstrations on actual quantum processors \cite{peruzzo2014,kandala2017} illustrate how the HL {\it ansatz} continues to inspire both classical and quantum computational methods. More recent works by Nakashima and Kurokawa \cite{nakashima2018,kurokawa2020}, as well as Sarwono \textit{et al.} \cite{sarwono2020}, show that HL-based ideas can still yield nearly exact potential energy curves for ground and excited states of H$_2$. 

In this work, we first revisit the main idea of the HL model by analytically deriving the ground-state energy of H$_2$ as a function of the nuclear separation for bonding and antibonding states. Next, we perform variational quantum Monte Carlo (VQMC) calculations using the HL wave function modified by a single variational parameter $\alpha$, which plays the role of an effective nuclear charge. We then propose a screening-modified HL model employing the same wave function as in the VQMC calculations. From these results, we construct an expression for $\alpha(R)$ as a function of $R$. Notably, this simple approach yields substantially improved agreement with the experimental bond length. We discuss and compare the dissociation energy, bond length, and vibrational frequency of the H$_2$ molecule in these three approaches.

\section{Hydrogen molecule in the HL model \label{sec:2}}

\begin{figure}[!ht]
\centering
\includegraphics[width=5in]{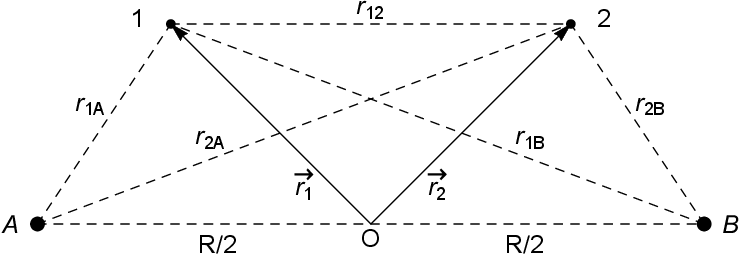}
\caption{Schematic of the hydrogen molecule showing electrons $i=1,\ 2$ and protons $j=A,\ B$. The electron-proton distances are $r_{ij}$. The electron-electron and proton-proton separations are $r_{12}$ and $R$, respectively.}
\label{Fig1}
\end{figure}

The hydrogen molecule comprises two protons, $A$ and $B$, and two electrons, $1$ and $2$, which interact via a pairwise Coulomb potential. Given the large mass difference between protons and electrons, the two particles move on different time scales. For this reason, we employ the Born-Oppenheimer approximation, which decouples the motions of electrons and protons, thus allowing the electronic Hamiltonian to be solved with the protons fixed at a distance $R$. The sum of the electronic and the proton-proton repulsion energies gives the total energy of the hydrogen molecule. Therefore, the Schr\"odinger equation for H$_2$ reads:
\begin{equation}
\hat{H}\,\Psi(\vec{r}_1,\vec{r}_2)=E_T\,\Psi(\vec{r}_1,\vec{r}_2).
\end{equation}
The corresponding non-relativistic many-body Hamiltonian in atomic units is
\begin{equation}
\hat{H}=-\frac{1}{2}{\nabla}_{1}^{2}-\frac{1}{2}{\nabla}_{2}^{2}-\frac{1}{r_{1A}}-\frac{1}{r_{1B}}-\frac{1}{r_{2A}}-\frac{1}{r_{2B}}+\frac{1}{r_{12}}+\frac{1}{R},
\label{ManyBodySchroedinger}
\end{equation}
where ${\nabla}_{i}^2$ is the Laplacian operator acting on the $i^{\text{th}}$ electronic coordinate, and \mbox{$r_{ij}=|\vec{r}_i-\vec{r}_j|$}. From left to right, the terms correspond to the kinetic energies of the electrons, the attractive potentials between electrons and protons, and the repulsive electron-electron and proton-proton potentials. Figure~\ref{Fig1} shows a geometrical representation of the hydrogen molecule, depicting all relevant distances.

Let us first consider the ground-state radial wave function for the $1s$ orbital of an isolated hydrogen atom, in which an electron $i$ is bound to a proton $j$. It is given by 
\begin{equation}
\phi(r_{ij})=\sqrt{\frac{1}{\pi}}\,e^{-r_{ij}},
\label{atomic1swf}
\end{equation}
The main idea of the HL model is to express the wave function of the H$_2$ molecule as a linear combination of products of atomic $1s$ orbitals $\phi(r_{ij})$, as follows:
\begin{equation}
\psi_{\pm}(\vec{r}_1,\vec{r}_2)=N_{\pm}\,
[\phi(r_{1A})\,\phi(r_{2B})\pm\phi(r_{1B})\,\phi(r_{2A})],
\label{HLwf}
\end{equation}
where each electron is bound to a different proton; that is, the electron $1$ (or $2$) and the proton $A$ (or $B$) form $1s$ bound states. Two possible wave functions arise due to the relative phase (sign $\pm$) between the product of $1s$ states, each associated with a corresponding normalization factor $N_{\pm}$. Since the wave function includes two electrons, Fermi-Dirac statistics must be applied to ensure the antisymmetry of the wave function $\Psi_{(s,s_z)}(\vec{r}_1,\vec{r}_2)$ under exchange of the electrons. The first allowed wave function is the singlet state:
\begin{equation}
\Psi_{(0,0)}(\vec{r}_1,\vec{r}_2)=\psi_+(\vec{r}_1,\vec{r}_2)\frac{1}{\sqrt{2}}(|\!\uparrow\downarrow\rangle-|\!\downarrow\uparrow\rangle),
\label{HLwfsinglet}
\end{equation}
where the subscript $s=s_1+s_2$ is the total spin angular momentum of the electrons, and $s_z=m_1+m_2$ is the sum of their projections along the $z$-axis. The remaining wave functions are the triplet states with $s=1$:
\begin{align}
\Psi_{(1,1)}(\vec{r}_1,\vec{r}_2)&=\psi_-(\vec{r}_1,\vec{r}_2)|\!\uparrow\uparrow\rangle, \nonumber \\ 
\Psi_{(1,0)}(\vec{r}_1,\vec{r}_2)&=\psi_-(\vec{r}_1,\vec{r}_2)\frac{1}{\sqrt{2}}(|\!\uparrow\downarrow\rangle+|\!\downarrow\uparrow\rangle), \nonumber \\ 
\Psi_{(1,-1)}(\vec{r}_1,\vec{r}_2)&=\psi_-(\vec{r}_1,\vec{r}_2)|\!\downarrow\downarrow\rangle.
\label{HLwftriplet}
\end{align}
It is important to note that the HL proposal becomes exact as the nuclear separation $R \to \infty$. By applying the normalization condition to the wave function
\begin{equation}
\int\int d^3r_1\,d^3r_2\,|\psi_{\pm}(\vec{r}_1,\vec{r}_2)|^2=\int\int d^3r_1\,d^3r_2\,\{N_{\pm}\,
[\phi(r_{1A})\,\phi(r_{2B})\pm\phi(r_{1B})\,\phi(r_{2A})]\}^2=1,
\label{normcondition}
\end{equation}
we obtain
\begin{equation}
N_\pm=\frac{1}{\sqrt{2\pm 2I_s^2}}.
\label{Nfactor}
\end{equation}
The integrals $I_s$ above are overlap integrals describing the superposition of atomic orbitals of an electron centered on different protons. These integrals take the following form
\begin{equation}
I_s=\int d^3r_i\,\phi(r_{ij})\,\phi(r_{ij'})=\frac{1}{\pi}\,\int d^3r_i\, e^{-(r_{ij}+r_{ij'})},
\label{Is}
\end{equation}
which yields
\begin{equation}
I_s=\left(1+R+\frac{R^2}{3}\right)e^{-R}.
\label{Is4}
\end{equation}
Substituting this result into Eq.~\eqref{Nfactor}, we find
\begin{equation}
N_\pm=\frac{1}{\sqrt{2}}\left[1\pm\left(1+R+\frac{R^2}{3}\right)^2e^{-2R}\right]^{-1/2}.
\label{Nfactor2}
\end{equation}
 
We can now calculate, term by term, the expectation value of the Hamiltonian in Eq.~\eqref{ManyBodySchroedinger}. Let us first consider the electronic kinetic energy
\begin{equation}
\langle \hat{K}_i\rangle=\left\langle-\frac{1}{2}\vec{\nabla}_{i}^{2}\right\rangle
=\int\int d^3r_1\,d^3{r_2}\, \psi_{\pm}^*(\vec{r}_1,\vec{r}_2)\left(-\frac{1}{2}\vec{\nabla}_{i}^{2}\right)\psi_{\pm}(\vec{r}_1,\vec{r}_2).
\label{KE} 
\end{equation}
By adding and subtracting the following terms within the integrands
\begin{align}
\langle \hat{K}_i\rangle
&=N_{\pm}\int\int d^3r_1\,d^3{r_2} \psi_{\pm}^*(\vec{r}_1,\vec{r}_2)\,\times\nonumber \\ 
&\left[\phi(r_{i'B})\left(-\frac{1}{2}\vec{\nabla}_{i}^{2}-\frac{1}{r_{iA}}+\frac{1}{r_{iA}}\right)\phi(r_{iA})
\pm\phi(r_{i'A})\left(-\frac{1}{2}\vec{\nabla}_{i}^{2}-\frac{1}{r_{iB}}+\frac{1}{r_{iB}}\right)\phi(r_{iB})\right],
\label{KE1}
\end{align}
where $i\neq i'$. Recalling the $1s$ eigenvalue $E_{1s}$ of the hydrogen atom and the normalization condition from Eq.~\eqref{normcondition}, we obtain
\begin{equation}
\langle \hat{K}_i\rangle=E_{1s}+N_{\pm}\int\int d^3r_1\,d^3{r_2}\, \psi_{\pm}^*(\vec{r}_1,\vec{r}_2)\left[\frac{1}{r_{iA}}\phi(r_{i'B})\,\phi(r_{iA})\pm\frac{1}{r_{iB}}\phi(r_{i'A})\,\phi(r_{iB})\right].
\label{KE3}
\end{equation}
Since $\phi(r_{ij})$ is a real function, we can reduce the number of terms by noting that the molecular wave function is an even function. Thus, we arrive at
\begin{equation}
\langle\hat{K}_i\rangle=E_{1s}+2N_{\pm}^2\left[\int d^3r_i\, \frac{\phi^2(r_{iA})}{r_{iA}}\pm \int d^3r_i \,\frac{\phi(r_{iA})\phi(r_{iB})}{r_{iB}}\int d^3r_{i'} \,\phi(r_{i'B})\,\phi(r_{i'A})\right],
\label{KE4}
\end{equation}
where the first integral yields $\left\langle 1/r \right\rangle=1$ in atomic units, the second is denoted $I_{t}$, and the last is the overlap integral $I_s$, as previously discussed. After inserting $N_{\pm}$ from Eq.~\eqref{Nfactor} and using $E_{1s}=-1/2$\,E$_h$, we have
\begin{equation}
\langle\hat{K}_i\rangle=E_{1s}\left(1-2\frac{1\pm I_tI_s}{1\pm I_s^2}\right),
\label{KE5}
\end{equation}
with $\langle \hat{K}_1\rangle=\langle \hat{K}_2\rangle$. The integral $I_t$ can be written as
\begin{equation}
I_t=\int d^3r_i \frac{\phi(r_{ij})\,\phi(r_{ij'})}{r_{ij}}=\frac{1}{\pi}\int d^3 r_i\,\frac{e^{-(r_{ij}+r_{ij'})}}{r_{ij}}=\left(1+{R}\right)e^{-R}.
\label{It2}
\end{equation}

The expectation value of the electron-proton attractive potential is
\begin{equation}
\langle V_{ij}\rangle=\left\langle-\frac{1}{r_{ij}}\right\rangle=-\int\int d^3r_1\,d^3{r_2}\, \psi_{\pm}^*(\vec{r}_1,\vec{r}_2)\,\frac{1}{r_{ij}}\,\psi_{\pm}(\vec{r}_1,\vec{r}_2).
\label{Vep}
\end{equation}
Following the same procedure that leads from Eq.~\eqref{KE4} to Eq.~\eqref{KE5}, we evaluate 
\begin{equation}
\langle V_{ij}\rangle=-N_{\pm}^2 \left[1\pm2\,I_{t}\,I_s+\int d^3r_i \left(\frac{\phi(r_{iB})^2}{r_{iA}}\right)\right],
\label{Vep2}
\end{equation}
where the remaining integral is $I_d$. After substituting Eq.~\eqref{Nfactor} and $E_{1s}$, we have
\begin{equation}
\langle V_{ij}\rangle=E_{1s} \left(\frac{1\pm2\,I_{t}\,I_s+I_d}{1\pm I_s^2}\right), 
\label{Vep3}
\end{equation}
and the integral $I_d$ is given by
\begin{equation}
I_d=\int d^3r_i\frac{\phi(r_{ij})^2}{r_{ij'}}=\frac{1}{\pi}\int d^3 r_i\frac{e^{-2r_{ij}}}{r_{ij'}}=\frac{1}{R}-\left(\frac{1}{R}+1\right)e^{-2R},
\label{Id4}
\end{equation}
where $j\neq j'$. The expectation value of the repulsive interaction between electrons is
\begin{equation}
\langle V_{12}\rangle=\left\langle\frac{1}{r_{12}}\right\rangle=\int\int d^3r_1\,d^3{r_2}\, \psi_{\pm}^*(\vec{r}_1,\vec{r}_2)\,\frac{1}{r_{12}}\,\psi_{\pm}(\vec{r}_1,\vec{r}_2),
\label{Vee}
\end{equation}
and it is composed of the following integrals
\begin{equation}
\langle V_{12}\rangle=2\,N_{\pm}^2\int\int d^3r_1\,d^3{r_2} \left[\frac{\phi(r_{1A})^2\,\phi(r_{2B})^2}{r_{12}}\pm\frac{\phi(r_{1A})\,\phi(r_{2B})\phi(r_{1B})\,\phi(r_{2A})}{r_{12}}\right],
\label{Vee2}
\end{equation}
referred to as the Coulomb integral $I_\textrm{Coul}$ and the exchange integral $I_\textrm{x}$, respectively. While $I_\textrm{Coul}$ can be interpreted as the classical Coulomb interaction between the electronic distributions $\phi(r_{1A})^2$ and $\phi(r_{2B})^2$, $I_\textrm{x}$ does not have any classical counterpart and comes from the antisymmetry of the wave function. Thus,
\begin{equation}
\langle V_{12}\rangle=-2\,E_{1s}\left(\frac{I_\textrm{Coul}\pm I_\textrm{x}}{1\pm I_s^2}\right),
\label{Vee3}
\end{equation}
where the integral $I_\textrm{Coul}$ is straightforward to calculate and yields
\begin{equation}
I_\textrm{Coul}=\frac{1}{R}-\frac{1}{R}\left(1+\frac{11R}{8}+\frac{3R^2}{4}+\frac{R^3}{6}\right)e^{-2R}.
\label{IC4}
\end{equation}
The integral $I_\textrm{x}$ is challenging to evaluate and is given by
\begin{equation}
I_\textrm{x}=\frac{1}{\pi^2}\int\int\,d^3r_1\,d^3r_2\,\frac{1}{r_{12}}e^{-(r_{1A}+r_{1B}+r_{2A}+r_{2B})}.
\label{IX}
\end{equation}
One way to handle this integral is by rewriting the spatial variables in prolate spheroidal coordinates. Assuming the positions of nuclei $A$ and $B$ as the foci positions of the coordinate system (note that $r_{iA}\pm r_{iB}$ represent the sum and the difference of distances to the foci), we can write
\begin{equation}
\xi_i=\frac{r_{iA}+r_{iB}}{R}, \quad\text{and}\quad
\eta_i=\frac{r_{iA}-r_{iB}}{R},
\label{pscoord}
\end{equation}
where $\xi_i=\cosh{\mu}$ and $\eta_i=\cos\nu$. The curves of constant $\xi_i\in[1,\infty)$ represent prolate spheroids; the curves of constant $\eta_i\in[-1,1]$ are hyperboloids of revolution; and $\phi_i\in[0,2\pi)$ is the azimuthal angle. We have in Cartesian coordinates,
\begin{align}
x_i&=\frac{R}{2}\sqrt{(\xi_i^2-1)(1-\eta_i^2)}\cos\phi,\nonumber \\ 
y_i&=\frac{R}{2}\sqrt{(\xi_i^2-1)(1-\eta_i^2)}\sin\phi,\nonumber \\ 
z_i&=\frac{R}{2}\xi_i\eta_i.
\label{CartCoord}
\end{align}
After taking the Jacobian, the infinitesimal volume element becomes
\begin{equation}
d^3r_i=-\frac{R^3}{8}(\xi_i^2-\eta_i^2)d\xi_id\eta_id\phi_i.
\label{psvolelem}
\end{equation}
After replacing Eqs.~\eqref{pscoord} and~\eqref{psvolelem} into Eq.~\eqref{IX} and solving the integral in $\phi_1$ and $\phi_2$, we arrive at
\begin{equation}
I_\textrm{x}=\frac{R^6}{16}\int_{-1}^1 \int_{1}^\infty \int_{-1}^1 \int_{1}^\infty d\xi_1\,d\eta_1\,d\xi_2\,d\eta_2 \left(\xi_1^2-\eta_1^2\right)\left(\xi_2^2-\eta_2^2\right)\frac{e^{-R(\xi_1+\xi_2)}}{r_{12}}.
\label{IX2}
\end{equation}
We now expand $1/r_{12}$ in terms of Legendre polynomials using the result from Y. Sugiura \cite{sugiura1927uber} and a reference therein \cite{neumann1887vorlesungen}. Then
\begin{equation}
\frac{1}{r_{12}}=\frac{1}{R}\sum_{k=0}^{\infty}(2k+1)\,P_k(\xi_<)\,Q_k(\xi_>)\,P_k(\eta_1)\,P_k(\eta_2),
\label{expansion}
\end{equation}
where $P_k(\xi_<)$ and $Q_k(\xi_>)$ are the Legendre polynomials of the first and second kinds, respectively. The arguments $\xi_<$ and $\xi_>$ stand for the smaller and larger values between $\xi_1$ and $\xi_2$, i.e., $(\xi_<,\xi_>)=(\xi_1,\xi_2)$ for $\xi_1<\xi_2$ and $(\xi_<,\xi_>)=(\xi_2,\xi_1)$ for $\xi_2<\xi_1$. After replacing Eq.~\eqref{expansion} in Eq.~\eqref{IX2}, it is not difficult to verify that for $k=1$ and $k>2$,
\begin{equation}
\int_{-1}^1 \,d\eta_i\,\left(\xi_i^2-\eta_i^2\right)P_k(\eta_i)=0,
\label{nullterms}
\end{equation}
whereas the remaining integrals for $k=0$ and $k=2$ are 
\begin{align}
\int_{-1}^1 \,d\eta_i\,\left(\xi_i^2-\eta_i^2\right)P_0(\eta_i)&=2\left(\xi_i^2-\frac{1}{3}\right),\nonumber \\ 
\int_{-1}^1 \,d\eta_i\,\left(\xi_i^2-\eta_i^2\right)P_2(\eta_i)&=-\frac{4}{15},
\label{nonnullterms}
\end{align}
and the respective Legendre polynomials are
\begin{align}
P_0(\xi_i)&=1,\nonumber \\
Q_0(\xi_i)&=\log{\left(\frac{\xi_i+1}{\xi_i-1}\right)},\nonumber \\
P_2(\xi_i)&=\frac{3\,\xi_i^2-1}{2},\nonumber \\
Q_2(\xi_i)&=-3\,\xi_i+P_2(\xi_i)\,Q_0(\xi_i).
\label{Legk0&k2}
\end{align}
After plugging all these results into Eq.~\eqref{IX2} and integrating it, we have
\begin{align}
I_\textrm{x}&=e^{-2R}\left[\frac{5}{8}-\frac{23R}{20}-\frac{3R^2}{5}-\frac{R^3}{15}\right] \nonumber\\
&+\frac{6I_s^2}{5R}\left[\gamma+\log{\left(R\right)}+\left(\frac{\bar{I}}{I_s}\right)^2\text{Ei}\left(-4R\right)-2\left(\frac{\bar{I}}{I_s}\right)\text{Ei}\left(-2R\right)\right],
\label{IX3}
\end{align}
where $\gamma\approx 0.5772$ is the Euler constant, $\text{Ei}(x)$ is the exponential integral function given by
\begin{equation}
\text{Ei}(x)=\int_{-x}^{\infty}dt\,\frac{e^{-t}}{t},
\label{Ei}
\end{equation}
and $\bar{I}$ is given as follows
\begin{equation}
\bar{I}=e^{R}\left[1-R+\frac{R^2}{3}\right].   
\label{Ibar}
\end{equation}

Finally, the expectation value of the proton-proton Coulomb potential is straightforward to compute and results in
\begin{equation}
\langle V_{AB}\rangle=-\frac{2\,E_{1s}}{R}.
\label{Vpp}
\end{equation}
Therefore, after collecting the results for all the Hamiltonian terms above, the expectation value of $\hat{H}$ is
\begin{equation}
\langle \hat{H} \rangle_\pm=
2\,E_{1s}\left(1+\frac{\pm2\,I_{t}\,I_s+2\,I_d-I_C\mp I_{X}}{1\pm I_s^2}-\frac{1}{R}\right)=E^{HL}_{T\pm}(R).
\label{H}
\end{equation}
The behavior of $E^{HL}_{T\pm}$ as a function of $R$ reveals some meaningful features of H$_2$. First, the $E^{HL}_{T+}$ curve has a minimum at a finite nuclear separation $R$, which correctly reflects the formation of a molecule. As a result, both the binding energy and equilibrium bond length can be determined in this case. On the other hand, $E^{HL}_{T-}$ predicts a minimum at $R\to\infty$, corresponding to the situation where we have two separated atoms, rather than a molecule. In this context, the choice $\psi_+$ is usually called {\it bonding} molecular orbital and $\psi_-$ the {\it antibonding} one. Second, as $R\to 0$ we have $E^{HL}_{T\pm}\to\infty$, due to repulsive forces. These forces arise from Coulomb repulsion and the Pauli exclusion principle on the electrons. Third, as $R\rightarrow\infty$, we get $E^{HL}_{T\pm}=-1.0$\,E$_h$, the energy of two independent hydrogen atoms. Although this model represents a rich molecular picture, the nuclear potential seen by each electron is constant regardless of the distance $R$. As a consequence, the width of the atomic orbitals also remains unchanged. In the following sections, we show how this model can be extended by incorporating a screened nuclear potential into the HL wave function.

\section{Variational quantum Monte Carlo}\label{sec:3}

The variational quantum principle is the conceptual starting point for introducing the variational quantum Monte Carlo method. Consider a trial wave function $|\Psi^{(\alpha)}_T\rangle$ with a variational parameter $\alpha$. We begin by recalling the Schr\"odinger equation,
\begin{equation}
\hat{H}|\Psi^{(\alpha)}_T\rangle=E^{(\alpha)}_T|\Psi^{(\alpha)}_T\rangle,
\label{Schrodvar}
\end{equation}
where the operator $\hat{H}$ applied to $|\Psi^{(\alpha)}_T\rangle$ yields the variational total energy $E^{(\alpha)}_T$. Since $E^{(\alpha)}_T$ is real, we can rewrite the previous expression as 
\begin{equation}
E^{(\alpha)}_T[\Psi^{(\alpha)}_T(\vec{r})]=\frac{\langle\Psi^{(\alpha)}_T|\hat{H}|\Psi^{(\alpha)}_T\rangle}{\langle\Psi^{(\alpha)}_T|\Psi^{(\alpha)}_T\rangle},
\label{Schrodvar2}
\end{equation}
where $E^{(\alpha)}_T[\Psi^{(\alpha)}_T(\vec{r})]$ is a functional of the unnormalized trial wave function $\Psi^{(\alpha)}_T$ and $\vec{r}=\{\vec{r}_1,\vec{r}_2, \dots, \vec{r}_N\}$ for brevity. The variational principle determines the stationary energy by varying the trial wave function with respect to the parameter $\alpha$. The functional $E^{(\alpha)}_T[\Psi^{(\alpha)}_T(\vec{r})]$ can be rewritten as 
\begin{equation}
E^{(\alpha)}_T[\Psi^{(\alpha)}_T(\vec{r})]=\frac{\int d\vec{r}\, \Psi^{(\alpha)}_T(\vec{r})\,\hat{H}\,\Psi^{(\alpha)}_T(\vec{r})}{\int d\vec{r}\, |\Psi^{(\alpha)}_T(\vec{r})|^2} = \frac{\int d\vec{r}\, |\Psi^{(\alpha)}_T(\vec{r})|^2\left(\frac{\hat{H}\,\Psi^{(\alpha)}_T(\vec{r})}{\Psi^{(\alpha)}_T(\vec{r})}\right)}{\int d\vec{r}\, |\Psi^{(\alpha)}_T(\vec{r})|^2},
\label{ETvar}
\end{equation}
where the local energy $E^{(\alpha)}_L(\vec{r})$ for a given $\alpha$,
\begin{equation}
E^{(\alpha)}_L(\vec{r})=\frac{\hat{H}\,\Psi^{(\alpha)}_T(\vec{r})}{\Psi^{(\alpha)}_T(\vec{r})},
\label{localenergy}
\end{equation}
is weighted by a probability density
\begin{equation}
p(\vec{r})=\frac{|\Psi^{(\alpha)}_T(\vec{r})|^2}{\int d\vec{r}\, |\Psi^{(\alpha)}_T(\vec{r})|^2}.
\label{probdensity}
\end{equation}
Eq.~\eqref{ETvar} can be evaluated numerically as follows:
\begin{equation}
E^{(\alpha)}_T=\frac{1}{N_s}\sum_{i=1}^{N_s} E^{(\alpha)}_L(\vec{r}_i),
\label{totalenergy}
\end{equation}
where $N_s$ corresponds to the number of samples. Therefore, we must sample $p(\vec{r})$ in such a way that it generates a sequence $\{\vec{r}_1,\vec{r}_2,\dots,\vec{r}_{N_s}\}$ where the frequency of each $\vec{r}_i$ is proportional to $p(\vec{r})$. The variance and standard error of $E^{(\alpha)}_T$ are given, respectively, by
\begin{equation}
\sigma^2=\frac{1}{N_s-1}\sum_{i=1}^{N_s}\left[E^{(\alpha)}_L(\vec{r}_i)-E^{(\alpha)}_T\right]^2 \quad\text{and}\quad \Delta E^{(\alpha)}_T=\frac{\sigma}{\sqrt{N_s}}.
\label{sigmaerror}
\end{equation}

The Monte Carlo approach for sampling the $N_s$ configurations of $E^{(\alpha)}_{L\pm}(\vec{r})$ is based on the Metropolis algorithm \cite{metropolis1953equation}. The idea behind the Metropolis algorithm is to generate a sequence of random samples, testing each one against a specified probability distribution function. The outcome of each test is either the acceptance or rejection of the new sample. If a new sample is accepted, it replaces the previous one; otherwise, the previous sample is retained for the next trial. Each sample is used to compute the local energy $E^{(\alpha)}_L(\vec{r})$. The frequency with which the values $E^{(\alpha)}_L(\vec{r})$ appear is proportional to the probability density $p(\vec{r})$, ensuring an accurate estimate of $E^{(\alpha)}_T$ after the sampling. 

\section{Screening in the HL model}\label{sec:4}

\begin{figure}[!ht]
\centering
\includegraphics[width=5in]{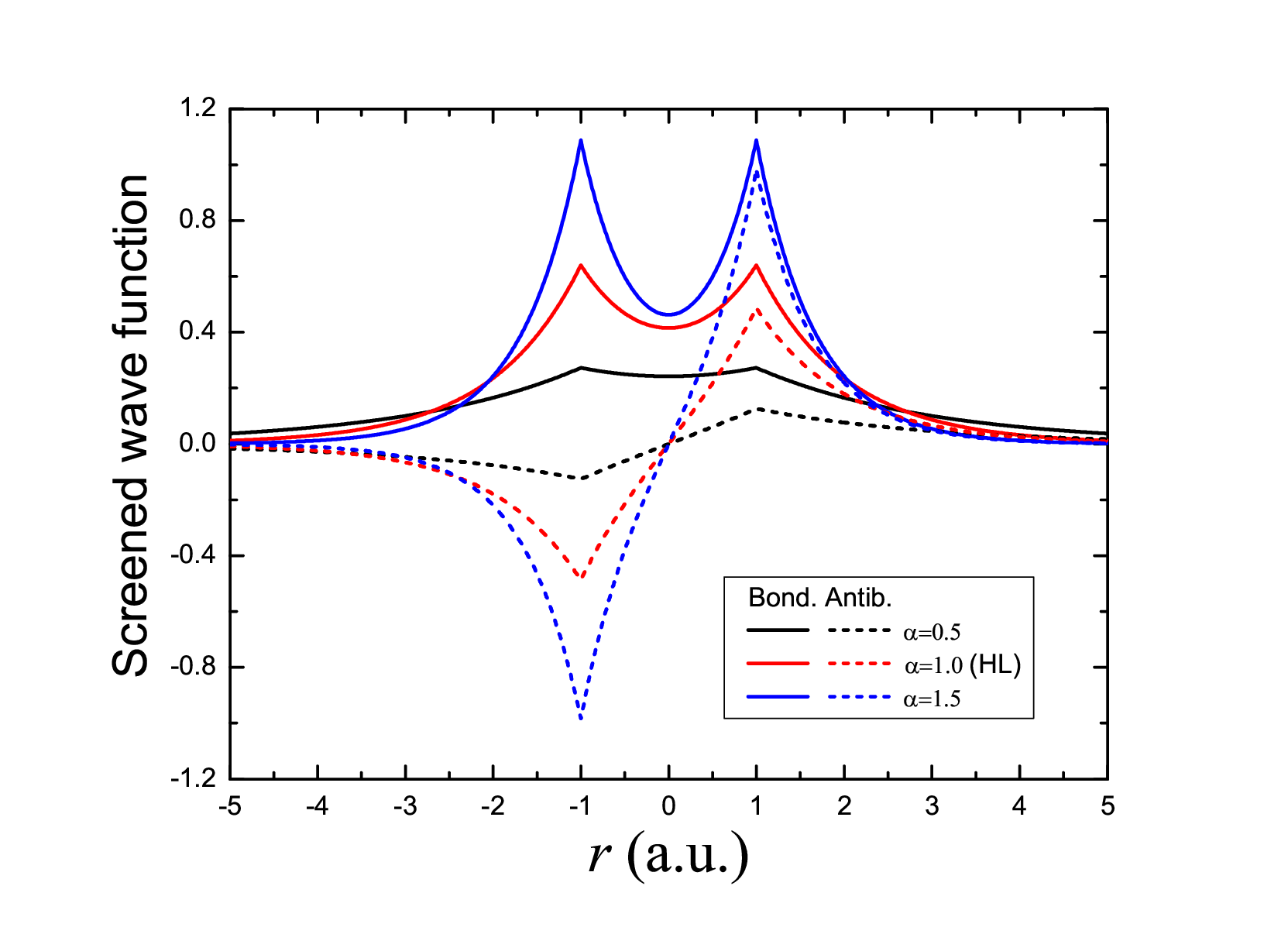}
\caption{Bonding (solid lines) and antibonding (dashed lines) screened wave functions for $\alpha=0.5$ (black), $1.0$ (red), and $1.5$ (blue), with the nuclear separation fixed at $R=2.0$. The functions are defined in Eq.~\eqref{HLwfT} and constructed using the atomic orbitals given in Eq.~\eqref{HLwfTalpha}. Note that $\alpha=1.0$ corresponds to the original HL wave function.}
\label{Fig2}
\end{figure}

Electrons behave as negatively charged spherical clouds that partially screen the electrostatic potential of the positive nuclei in the hydrogen molecule. This means that each electron experiences an effective nuclear charge from its nearby proton, which decreases as the other electron approaches. This fact provides a clue about how to incorporate a screening potential into the HL model. Let us introduce a parameter $\alpha$ that controls the width of all atomic orbitals, effectively representing the screened nuclear charge experienced by each electron. This parameter can be added by rewriting the atomic orbital as
\begin{equation}
\phi^{(\alpha)}(r_{ij})=\sqrt{\frac{\alpha^3}{\pi }}\, e^{-\alpha\,{r_{ij}}}. 
\label{HLwfTalpha}
\end{equation}
Thus, a trial wave function based on the HL {\it ansatz} can be written as follows
\begin{equation}
\begin{split}
\Psi^{(\alpha)}_{T \pm}(\vec{r}_1,\vec{r}_2)&=\phi^{(\alpha)}(r_{1A})\phi^{(\alpha)}(r_{2B})\pm\phi^{(\alpha)}(r_{1B})\phi^{(\alpha)}(r_{2A}) \\
&=e^{-{\alpha}\,(r_{1A}+r_{2B})}\pm e^{-{\alpha}\,(r_{1B}+r_{2A})}.
\end{split}
\label{HLwfT}
\end{equation}
Note that both the local energy in Eq.~\eqref{localenergy} and the probability density in Eq.~\eqref{probdensity} depend on the ratio of wave functions. Thus, the normalization factor can be neglected in this context. As shown in Fig.~\ref{Fig2}, increasing $\alpha$ leads to greater localization, and decreasing it results in delocalization. As the next step, this effective nuclear charge $\alpha$ will be chosen as the variational parameter in our Monte Carlo calculations. The Monte Carlo method identifies the optimal screening potential as a function of $R$ by varying the effective nuclear charge. This insight is useful for understanding how atomic orbitals dynamically evolve during bond formation or dissociation. Ultimately, this screened wave function could be employed to model such physical scenarios. Once we have this wave function, the local energy calculated from Eq.~\eqref{localenergy} is
\begin{align}
E^{(\alpha)}_{L\pm}(\vec{r})
&=-\frac{\alpha}{2}\left[\sum_{\substack{i,i'=1 \\ i\neq i'}}^2
\frac{e^{-{\alpha}\,(r_{iA}+r_{i'B})}\left({\alpha}-\frac{2}{r_{iA}}\right)\pm
e^{-{\alpha}\,(r_{iB}+r_{i'A})}\left({\alpha}-\frac{2}{r_{iB}}\right)}{e^{-{\alpha}\,(r_{iA}+r_{i'B})}\pm e^{-{\alpha}\,(r_{iB}+r_{i'A})}}\right]\nonumber\\
&\times \left[-\frac{1}{r_{1A}}-\frac{1}{r_{2A}}-\frac{1}{r_{1B}}-\frac{1}{r_{2B}}+\frac{1}{r_{12}}+\frac{1}{R}\right].
\label{localenergynum}
\end{align}

It is worth noting that a few careful numerical choices must be made to ensure proper Monte Carlo sampling. The procedure described above must be performed by varying $\alpha$ until the optimal value $\alpha_0$ is found, which corresponds to the stationary total energy $E^{(\alpha_0)}_T$ for each proton-proton distance $R$. These calculations must employ a sufficiently large sampling size, such as $N_s=10^8$ (computed after reaching equilibrium), to ensure smooth energy curves with well-defined minima. Finally, a displacement parameter $\delta(\alpha)$ was adjusted to achieve a sample acceptance rate of approximately $50\%$ in all Metropolis tests. After extensive trials, we found a power-law relation given by $\delta(\alpha)=d/\alpha$, where $d=1.45$ and $1.75$ for the antibonding and bonding states, respectively. The calculation of $E^{(\alpha_0)}_T$ over a wide range of $R$ values yields a set of points that defines an optimized stochastic curve, $E^{(\alpha_0)}_T(R)$, for H$_2$. This curve can then be compared with the original HL model and our proposed approach, as discussed in the following paragraphs. 

\begin{figure}[!ht]
\centering
\includegraphics[width=5in]{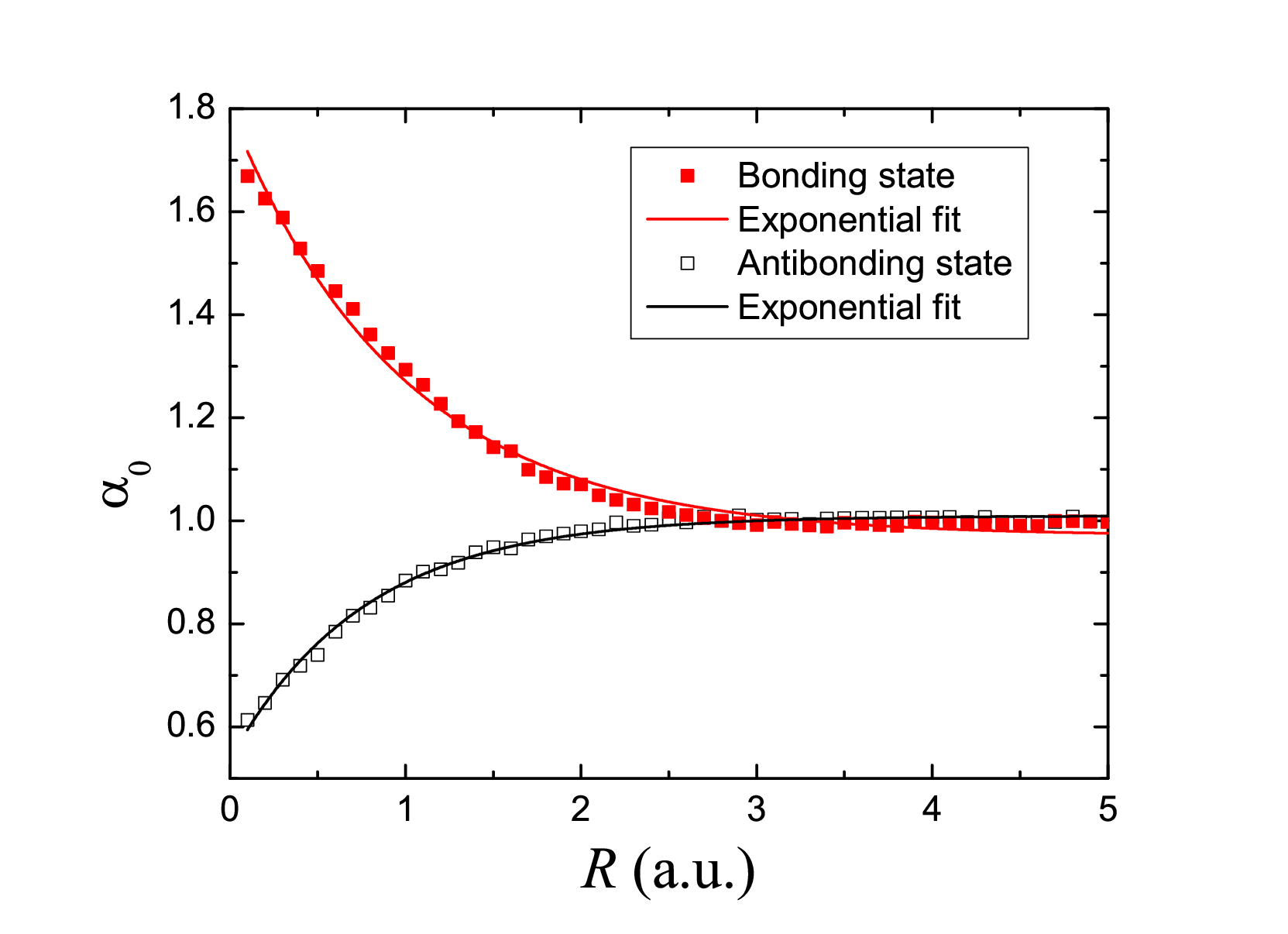}
\caption{Exponential fits of $\alpha_{0\pm}(R)$ as defined in Eq.~(\ref{Qeff}): $\alpha_{0+}(R)=0.970(5)+0.826(13)e^{-1.01(3)R}$ for the bonding state (red line) and $\alpha_{0-}(R)=1.01(0)-0.473(7)e^{-1.30(3)R}$ for the antibonding state (black line). In both cases, the fitted functions yield $\alpha_{0\pm}(R)\approx 1$ as $R\to \infty$, recovering the original HL model without screening. For $R\to 0$, the effective charges approach $1.80$ and $0.54$ for the bonding and antibonding states, respectively. The red solid and black open square points were obtained from VQMC calculations with a sampling size of $N_s=10^8$.}
\label{Fig3}
\end{figure}

In Fig.~\ref{Fig3}, the behavior of the optimized effective nuclear charge $\alpha_0$ as a function of $R$ is shown. For both the bonding and antibonding states, as $R\to\infty$, the system approaches two isolated hydrogen atoms, where the effective charge seen by each electron corresponds to that of an unscreened proton. In this limit, $\alpha_0\to 1$, and the HL approach becomes exact. Conversely, as the two bonding electrons approach one another ($R\to 0$), the orbital width should decrease (i.e., $\alpha_0$ increases) to reduce the extensive overlap between their orbitals. In this limit, both protons collapse into a single nucleus, and the system resembles a helium atom in the bonding state. Using a variational approach with a wave function defined as a product of atomic orbitals, it is well known that the effective charge seen by each electron in helium is $\alpha_{\text{He}}=27/16\approx 1.69$ \cite{clementi1963,griffiths1995}. For the antibonding state, on the other hand, the electrons’ orbitals spread out as $R\to 0$ (i.e., $\alpha_0$ decreases). This occurs due to destructive interference, which pushes the orbitals away from the region between the nuclei. As a result, the effective charge seen by each electron becomes very small—smaller than the charge of a single proton. In other words, the antibonding state reduces the effective nuclear charge due to the opposite phases of the electrons’ wave functions.

To find an analytical expression for $\alpha_0(R)$ in the interval between $\alpha_{\text{He}}$ and $1.0$, we consider protons $A$ and $B$ as positive point charges partially surrounded by negatively charged spherical volumes representing electrons $1$ and $2$ (see Fig.~\ref{Fig1}). All charges are assumed to have the same magnitude. From the perspective of electron $1$, the charge distribution of electron $2$ is given by $\rho=|\phi^{(\alpha)}(r)|^2$, where $\phi^{(\alpha)}(r)$ is defined in Eq.~\eqref{HLwfTalpha}. Inspired by the form given by the integral of Eq.~\eqref{Id4}, we propose a Coulomb potential of the form $\alpha_0(R)/R$ experienced by one electron in the bonding state, where the effective charge is 
\begin{equation}
\alpha_{0\pm}(R)=\beta_\pm+\left(\alpha_{\text{He}}-\gamma_\pm\right)e^{-\lambda_\pm R}.
\label{Qeff}
\end{equation}
As shown in Fig.~\ref{Fig3}, both $\alpha_0$ curves (bonding and antibonding) are fitted using the exponential function given above. For the bonding state, the best-fit parameters are \mbox{$\beta_+ = 0.970(5)$}, \mbox{$\alpha_{\text{He}}-\gamma_+ =0.826(13)$} and \mbox{$\lambda_+=1.01(3)$}, whereas for the antibonding state they are \mbox{$\beta_- = 1.01(0)$}, \mbox{$\alpha_{\text{He}}-\gamma_- =-0.473(7)$} and \mbox{$\lambda_-=1.30(3)$}. Note that in both cases, $\alpha_{0\pm}\approx 1$ as $R\to\infty$ corresponding to the unscreened limit. On the other hand, for $R\to 0$ in the bonding state, $\alpha_{0\, +}$ deviates by about $6.5\%$ from $\alpha_{\text{He}}$. 

\begin{figure}[!ht]
\centering
\includegraphics[width=5in]{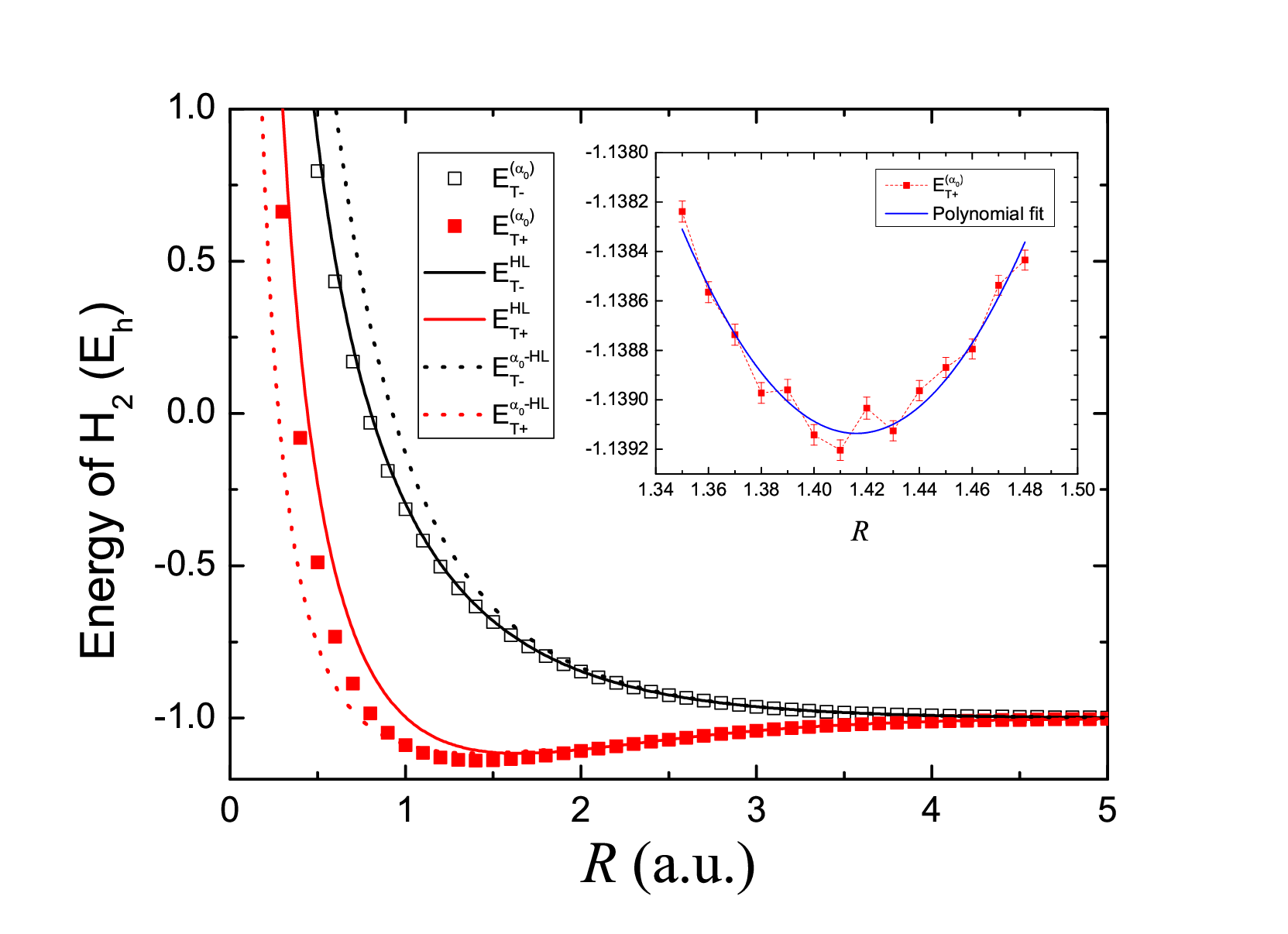}
\caption{Total energy of the hydrogen molecule as a function of the inter-proton distance $R$. The solid black and red lines correspond to the HL curves for the antibonding and bonding states, respectively, as given by Eq.~\eqref{H}. The black open and red solid square points were obtained from VQMC calculations with a sampling size of $N_s=10^8$. The dotted black and red lines represent the curves from our $\alpha_0$-HL proposal, constructed by incorporating the exponential fits of $\alpha_{0\pm}(R)$ into the HL model. The inset shows the bonding-state VQMC curve $E(R)$ near the energy minimum, based on 14 sampled points, along with the corresponding quadratic fit: $E(R)=-0.760(24)-0.536(35)R+0.189(12)R^2$.}
\label{Fig4}
\end{figure}

Based on Eq.~\eqref{Qeff} and the set of parameters described in the previous paragraph, we propose a modification to the HL model by incorporating $\alpha_{0\pm}$ into the wave function. This approach is hereafter referred to as the $\alpha_0$-HL model. Figure~\ref{Fig4} displays the energy of H$_2$ as a function of $R$, comparing the original HL model, VQMC calculations, and the $\alpha_0$-HL approach for both bonding and antibonding states. The VQMC-optimized curves yield lower energies than their analytical counterparts. This indicates that during bond formation or molecular dissociation, optimizing the widths of the atomic orbitals balances attractive and repulsive forces, thereby minimizing the total energy. For the bonding state, the HL model yields $E_0=-1.12$ E$_h$ at $R_0=1.64$ a.u., whereas the VQMC calculation gives $E_0=-1.14$ E$_h$ at $R_0=1.42$ a.u. Since the $\alpha_0$-HL model involves only a rescaling of distances, it yields the same energy $E_0$ as the HL model, but with a bond length of $R_0=1.40$ a.u., in exact agreement with the experimental value. 

Next, we examine the vibrational frequency of H$_2$. The inset of Fig.~\ref{Fig4} shows a polynomial fit to the energy of H$_2$ around its global minimum. The vibrational frequency $\nu_0$ of H$_2$ can be calculated using
\begin{equation}
\nu_0=\frac{1}{2\pi c} \sqrt{\frac{k}{\mu}},
\end{equation}
where $c$ is the speed of light, $\mu$ is the reduced mass of the hydrogen molecule, and $k=\partial^2 E/\partial R^2$ around the minimum. The HL model yields $\nu_0=3811$ cm$^{-1}$, while the $\alpha_0$-HL approach gives $\nu_0=3381$ cm$^{-1}$. Both deviate in the opposite direction from the VQMC result ($\nu_0=4471$ cm$^{-1}$) and the experimental value ($\nu_0=4380$ cm$^{-1}$).

\begin{figure}[!ht]
\centering
\includegraphics[width=5in]{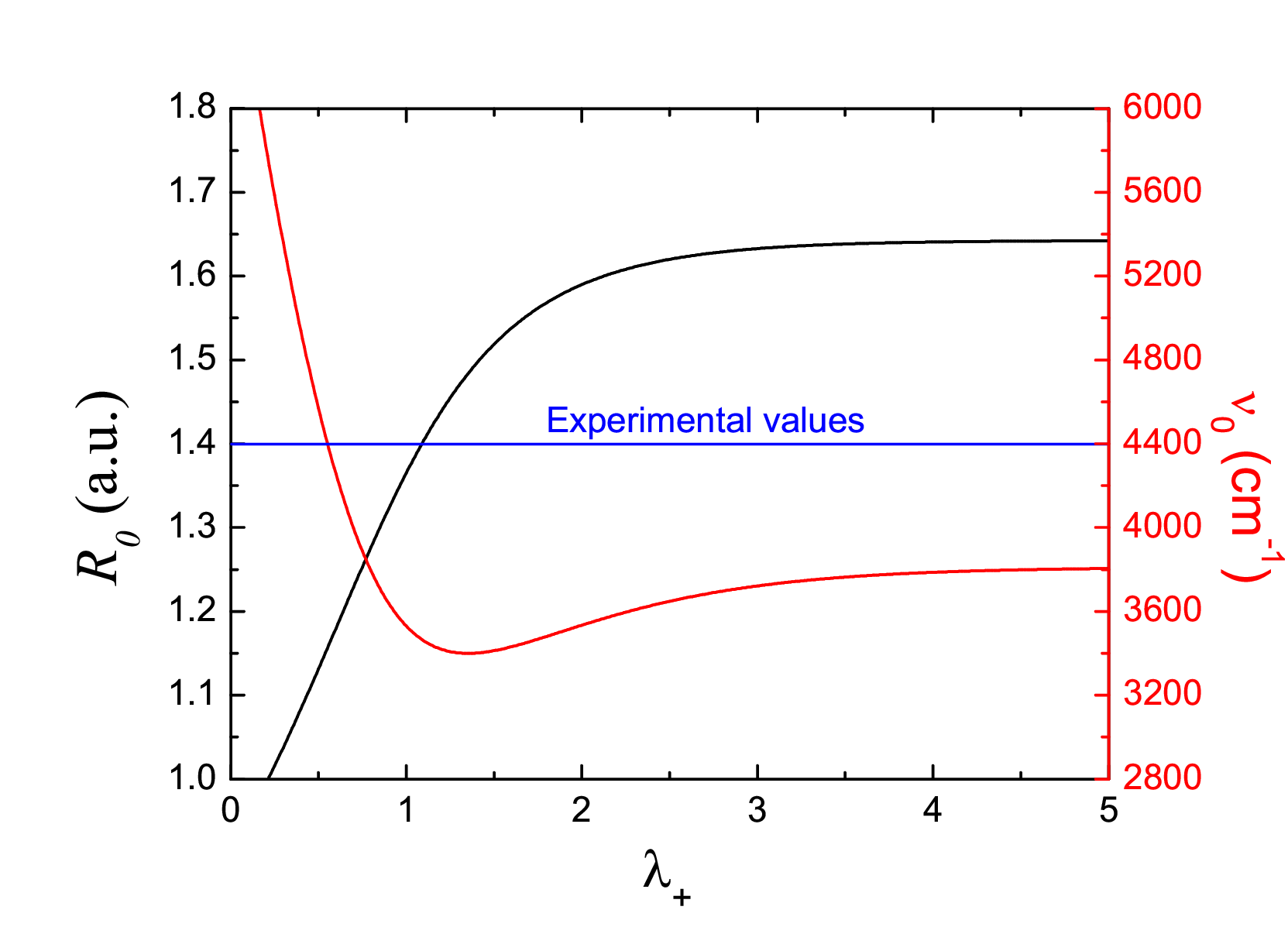}
\caption{Bond length (black) and vibrational frequency (red) of the hydrogen molecule as functions of the parameter $\lambda_+$ in Eq.~\eqref{Qeff}. The blue line indicates the corresponding experimental values.}
\label{Fig5}
\end{figure}

Let us now assume $\lambda_+$ to be the only free parameter in Eq.~\eqref{Qeff}, with $\beta_+= \gamma_+ = 1$, so that $\alpha_{0+}$ attains the exact values for $R\to 0$ and $R\to \infty$. In this case, the HL model is recovered when $\lambda_+ \gg 1$, since $\alpha_{0+} \to 1$. We can then examine how the bond length and vibrational frequency vary as functions of $\lambda_+$. The results are displayed in Fig.~\ref{Fig5}. As shown, in this simple model with only one free parameter, a single value of $\lambda_+$ is not sufficient to simultaneously reproduce both the experimental bond length and vibrational frequency of the H$_2$ molecule. While the expected $R_0$ is obtained with $\lambda_+^{R_0} \approx 1.09$, the correct value of $\nu_0$ curiously requires $\lambda_+^{\nu_0} \approx \lambda_+^{R_0}/2$. Therefore, this approach may be useful in simulations where $R_0$ or $\nu_0$ must be modeled independently, allowing $\lambda_+$ to be tuned to match experimental values. 
 
\section{Conclusion}\label{sec:5}

This paper revisits one of the earliest approaches to describing molecular systems within the Schr\"odinger formalism of quantum mechanics: the Heitler–London (HL) model. We present two methods for calculating ground-state properties of the
hydrogen molecule using the HL model and the variational quantum Monte Carlo (VQMC) approach. We show that bond formation and dissociation in H$_2$ affect the effective nuclear charge experienced by each electron and the shape of the
atomic orbitals as the inter-proton distance $R$ varies. We propose a simple model to incorporate screening into the HL wave function. This model can be used to describe the bond length $R_0$ and vibrational frequency $\nu_0$ of H$_2$.  

It is important to place our results in the broader context of the past decades of research on the hydrogen molecule. State-of-the-art variational and \emph{ab initio} calculations, including relativistic and quantum electrodynamical (QED) corrections, are capable of achieving spectroscopic accuracy for bond lengths, dissociation energies, and vibrational frequencies \cite{sims2006,pachucki2010,pachucki2013,puchalski2023}. Furthermore, modern computational schemes such as quantum Monte Carlo and explicitly correlated wave functions can essentially solve the H$_2$ problem to within experimental precision. More recently, hybrid classical–quantum algorithms \cite{peruzzo2014,kandala2017} have demonstrated that the hydrogen molecule continues to serve as a benchmark for quantum simulation platforms. In this context, the contribution of the present work is not to rival such high-precision techniques, but to provide a mathematically simple extension of the HL model. By introducing a simple screening parameter into the original HL wave function, we preserve the main ideas of the early quantum-mechanical treatment while capturing physically relevant effects absent in the 1927 model. We hope that our findings contribute to the development of improved---yet mathematically simple---variational wave functions that offer analytical insight into the mechanisms of molecular dissociation and bond formation, illustrating how ideas from the early days of quantum mechanics can be fruitfully reinterpreted in light of modern developments.

\section*{Acknowledgments}
This study was financed in part by the Coordena\c{c}\~ao de Aperfei\c{c}oamento de Pessoal de N\'ivel Superior - Brasil (CAPES) - Finance Code 001. It was also supported by Fundação de Amparo à Pesquisa Científica e Tecnológica de Santa Catarina (FAPESC), by the National Council for Scientific and Technological Development (CNPq) under Grants No. 409673/2022-6 and No. 309862/2021-3, and by the National Institute for Science and Technology of Quantum Information (INCT-IQ) under Grant No. \mbox{465469/2014-0}. E. P. M. Amorim thanks J. Longo for her careful reading of the manuscript and S. A. S. Vitiello for introducing the intricacies of the variational quantum Monte Carlo method.

\end{document}